\documentclass[useAMS,usenatbib]{mn2e} 
\usepackage{amsmath,graphicx}
\usepackage[usenames]{color}

\title[Mode identification in V2367\,Cyg]
{Mode identification in the high-amplitude $\delta$\,Scuti star
V2367\,Cyg}
\author[Ulusoy et al.] 	
{C. Ulusoy$^{1,2}$, T. G\"{u}lmez$^{1}$, I. Stateva$^{3}$, D. Dimitrov$^{3}$, I. Kh. Iliev$^{3}$, 
\newauthor
H. A. Kobulnicky$^{4}$, B. Ya\c{s}arsoy$^{5,6}$, M. \'{A}lvarez$^{7}$, R. Michel$^{7}$, 
\newauthor
L. Fox Machado$^{7}$, M. Damasso$^{8,9}$, A. Carbognani$^{8}$, D. Cenadelli$^{8}$, 
\newauthor
T. E. Pickering$^{10,13}$, J. Kwon$^{11,12}$, M. Tamura$^{11,12}$,  L. A. Balona$^{13}$\\
\\
$^1$Department of Physics, University of Johannesburg,  P.O. Box 524, APK Campus, 2006, Johannesburg, South Africa\\
$^2$Department of Physics, Izmir Institute of Technology, 35430, Izmir, Turkey\\
$^3$Institute of Astronomy with NAO, Bulgarian Academy of Sciences,
blvd.Tsarigradsko chaussee 72, Sofia 1784, Bulgaria\\
$^4$Department of Physics and Astronomy, University of Wyoming, P.O. Box
3905, Laramie, WY 82072, USA\\
$^5$Ege University Observatory, 35100, Bornova, Izmir, Turkey\\
$^6$In\"{o}n\"{u} University, Faculty of Arts and Sciences, Astronomy and Space Sciences Department, 44069, Malatya, Turkey\\
$^7$Observatorio Astron\'{o}mico Nacional, Instituto de Astronom\'{i}a, Universidad Nacional Aut\'{o}noma de M\'{e}xico,\\
Apartado Postal 877, 22830, Ensenada, B.C., M\'{e}xico\\
$^8$Astronomical Observatory of the Autonomous Region of the Aosta Valley (OAVdA) Loc. Lignan 39, 11020 Nus (Aosta), Italy\\
$^9$Department of Physics and Astronomy, University of Padova, vicolo dell'Osservatorio 3, I-35122 Padova, Italy\\
$^{10}$Southern African Large Telescope Foundation, P.O. Box 9, 7935 Observatory, Cape Town, South Africa\\
$^{11}$National Astronomical Observatory of Japan, 2-21-1 Osawa, Mitaka, Tokyo 181-8588, Japan\\
$^{12}$Department of Astronomical Science, Graduate University for Advanced Studies (Sokendai),\\
 2-21-1 Osawa, Mitaka, Tokyo 181-8588, Japan \\
$^{13}$South African Astronomical Observatory, P.O. Box 9, Observatory 7935,  Cape Town, South Africa \\
}

\begin{document}

\date{Accepted .... Received ...}

\pagerange{\pageref{firstpage}--\pageref{lastpage}} \pubyear{2011}

\maketitle

\label{firstpage}

\begin{abstract}
We report on a multi-site photometric campaign on the high-amplitude 
$\delta$\,Scuti star V2367\,Cyg in order to determine the pulsation modes.  
We also used high-dispersion spectroscopy to estimate the stellar parameters 
and projected rotational velocity.  Time series multicolour photometry was 
obtained during a 98-d interval from five different sites.  These data were 
used together with model atmospheres and non-adiabatic pulsation models to 
identify the spherical harmonic degree of the three independent frequencies 
of highest amplitude as well as the first two harmonics of the dominant
mode.  This was accomplished by matching the observed relative light 
amplitudes and phases in different wavebands with those computed by the 
models.  In general, our results support the assumed mode identifications in
a previous analysis of {\it Kepler} data.
\end{abstract}


\section{Introduction}

The high-amplitude delta Scuti (HADS) stars are defined as $\delta$\,Sct stars
with peak-to-peak light amplitudes in excess of about 0.3\,mag.  This
amplitude limit is an arbitrary one and there is no other physically
identifiably distinction between HADS and other $\delta$\,Sct stars.  The 
HADS stars rotate more slowly than other $\delta$~Sct stars with projected
rotationally velocities $v \sin i < 30$\,km\,s$^{-1}$.  They pulsate in one
or two dominant modes which are generally assumed to be radial modes, but 
when observed with greater precision from space many more modes are visible.  
A large number of combination frequencies and harmonics involving the modes 
of highest amplitude are usually present.  The pulsational behaviour of HADS
stars seems to be intermediate between that of multimode $\delta$~Scuti stars 
and Cepheids \citep{Breger2007}.  In fact, first-overtone classical Cepheids 
and HADS stars follow the same period -- luminosity relation with no 
discontinuity \citep{Soszynski2008}. 

The reason why $\delta$~Sct stars pulsate in many nonradial modes while the
more massive Cepheids pulsate only in radial modes is due to the behaviour
of nonradial pulsation modes in the stellar interior.  In the deep interior of
an evolved giant star such as a HADS star or low-mass Cepheid, nonradial p 
modes behave like high-order g modes.  As the stellar mass increases, these 
modes develop an increasingly large number of spatial oscillations which
leads to increased radiative damping.  Therefore we find fewer and fewer
nonradial modes in HADS stars of increasing mass until only radial modes
are present in what we then call a Cepheid \citep{Dziembowski1977}.  Indeed,
observations suggest that there is a group of HADS which pulsate just in the 
fundamental and first overtone modes with no nonradial modes 
\citep{Poretti2005} (they might be called low-mass Cepheids).   While
this seems to be the main reason why Cepheids do not pulsate in nonradial
modes, it may not be the only reason and the problem needs further 
investigation \citep{Mulet2007}.

The study of HADS stars is important because it is an interesting laboratory 
for investigating mode interaction and nonlinear behaviour as exemplified by 
the presence of combination frequencies.  The use of combination frequencies 
for mode identification is well-known in pure g-mode pulsators such as the
ZZ~Ceti stars \citep{Wu2001, Yeates2005, Montgomery2005}. \citet{Balona2012b} 
has recently investigated the problem in pulsating main sequence stars.  While
in ZZ~Ceti stars the combination frequencies arise from interactions in a 
convective zone in which certain approximations are valid, this is not the 
case in HADS and it turns out that combination frequencies cannot be used for 
mode identification.

Mode identification is crucial for extracting stellar parameters from the
pulsations.  The assumption that the modes of highest amplitude in HADS are
radial is based on the fact that when two dominant modes are present, their
period ratio often agrees with the expected period ratio for first overtone
to fundamental radial mode, $P_1/P_0$, which is in the range 0.77--0.78. 
The Petersen diagram \citep{Petersen1973}, which is a plot of $P_1/P_0$ as a 
function of $P_0$, is very useful for diagnostic purposes.  This ratio varies 
due to rotation with chemical composition \citep{Poretti2005} and by coupling
with the nearby quadrupole mode \citep{Suarez2006, Suarez2007}.  The calculated 
period ratio is also sensitive to the opacity \citep{Lenz2008}.

\begin{table}
\caption{Information on the photometric multisite campaign.  EUO: Ege University 
Observatory; BNAO-Rozhen: Bulgarian National Astronomical Observatory-Rozhen;  
OAVdA: The Astronomical Observatory of the Autonomous Region of the Aosta 
Valley; SPM: Observatorio de San Pedro M\'{a}rt\'{i}r;  RBO: Red Butte Observatory.  
Telescope apertures are in m.  The total number of nights observed and the 
start end ending Julian dates of the observations relative to JD\,24455000 
are given.}
\label{tab:log}
\begin{tabular}{llll}
\hline
Observatory & 
Site & 
Tel. &
CCD
\\
\hline
BNAO-Rozhen & Bulgaria & 0.60 & FLI PL9000          \\  
OAVdA       & Italy    & 0.81 & FLI  PL3041-1-BB    \\
SPM         & Mexico   & 0.84 & FLI Fairchild F3041 \\
EUO         & Turkey   & 0.40 & Apogee Alta-U47     \\
RBO         & USA      & 0.60 & Apogee Alta E47-UV  \\
\\
\hline
Observatory & 
Filters &
Dates &
Nights \\
\hline
BNAO-Rozhen & ${\rm UBVRI}$     &  2 & 736.34--737.54  \\
OAVdA       & ${\rm BVRI}$      &  3 & 737.37--739.56  \\
SPM         & ${\rm UBVR_cI_c}$ &  6 & 718.76--732.98  \\
EUO         & ${\rm BVRI}$      & 17 & 736.29--816.52  \\
RBO         & ${\rm UBVRI}$      &  1 & 741.68--741.93  \\
\hline
\end{tabular}
\end{table}

V2367\,Cyg (KIC\,9408694) was discovered in a ROTSE survey \citep{Akerlof2000} 
and confirmed as a HADS star by \citet{Jin2003} and \citet{Pigulski2009}. The 
star has been observed by the {\it Kepler} satellite and recently analyzed
by \citet{Balona2012a}.  The light curve is dominated by a single frequency
$f_1 = 5.6611$\,d$^{-1}$ of high amplitude. Two other independent modes with 
$f_2 = 7.1490$\,d$^{-1}$  and $f_3 = 7.7756$\,d$^{-1}$ have amplitudes an 
order of magnitude smaller than $f_1$. Nearly all the light variation is due 
to these three modes, harmonics of $f_1$ and combination frequencies with
$f_1$, although several hundred other frequencies of very low amplitude are 
also present in the {\it Kepler} photometry.

The star has a much higher projected rotational velocity ($v \sin i = 100 \pm 
10$\,km\,s$^{-1}$, \citet{Balona2012a}) than other HADS stars, which makes it 
unusual.  Such a large rotational velocity complicates the analysis and 
increases the likelihood of mode interaction.  The ratio $f_1/f_2 = 0.792$ is 
within the expected ratio of second- to first-overtone radial periods which 
is in the range 0.79--0.80. \citet{Balona2012a} attempted to model the 
observed frequencies as radial modes without mode interaction, but were not 
successful.  A model with mode interaction was found in which $f_1$ is 
identified as the fundamental radial mode and $f_2$ the first overtone radial 
mode coupled with a nearby quadrupole mode.

Considering the complex nature of the pulsation, the question arises as to 
whether the modes $f_1$ and $f_2$ are indeed radial and coupled modes as
suggested by the models.  The model proposed by \citet{Balona2012a} would be 
greatly strengthened if independent mode identification could be made.  Since 
the star is sufficiently bright and of high amplitude, multicolour photometry
can be used to provide this verification.  The purpose of this paper is to
describe a multi-site photometric and spectroscopic campaign on V2367\,Cyg and
subsequent analysis of mode identification.

\section{Spectroscopy}
The spectra were used to estimate the stellar atmospheric parameters and the 
projected rotational velocity, $v\,\sin\,i$. 
The spectroscopic observations were carried out with two instruments.
The Coud\`e spectra were obtained with the 2-m RCC telescope of the Bulgarian 
National Astronomical Observatory, Rozhen.  We observed the star during the
night of 2011 July 27 in three spectral regions: 4770--4970\,\AA, 
4400--4600\,\AA, and 5090--5290\,\AA.  These three regions contain the 
H$\beta$ line, Mg\,II $\lambda$\,4481\,\AA\ and a region rich in Fe lines.  
The instrument uses a Photometrics AT200 camera with a SITe SI003AB $1024 
\times 1024$ CCD, ($24\,\mu {\rm m}$ pixels) giving a typical resolution 
R = 16\,000 and S/N ratio  of about 40.  The exposure time was 1200\,s. 
The instrumental profile was checked by using the arc spectrum giving a full 
width at half maximum (FWHM) of about 0.4\,\AA. {\tt IRAF} standard procedures 
were used for bias subtracting, flat-fielding and wavelength calibration. The 
final spectra were corrected to the heliocentric wavelengths.

We also obtained spectra with the WIRO longslit spectrograph which uses an 
e2V 2048$^2$ CCD as the detector.  An 1800\,lines\,mm$^{-1}$ grating in first 
order yielded a spectral resolution of 1.5\,\AA\ near 5800\,\AA\ with a 
1.2\arcsec $\times$ 100\arcsec slit. The spectral coverage was 5250--6750~\AA.
Individual exposure times were 600\,s. Reductions followed standard longslit 
techniques. Multiple exposures were combined yielding final S/N ratios 
typically in excess of 60 near 5800\,\AA. Final spectra were Doppler corrected
to the heliocentric frame.  Each spectrum was then shifted by a small 
additional amount in velocity so that the Na\,I\,D $\lambda\lambda$\,5890, 
5996 lines were registered with the mean Na\,I line wavelength across the 
ensemble of observations. This zero-point correction to each observation is 
needed to account for effects of image wander in the dispersion direction when
the stellar FWHM of the point spread function was appreciably less than the 
slit width.  Because of these inevitable slit-placement effects on the 
resulting wavelength solutions at the level of $\leq$10\,km\,s$^{-1}$, radial 
velocity standards were not routinely  taken.

\section{Atmospheric parameters}
Model atmospheres were calculated using the {\tt ATLAS\,12} code. 
The {\tt VALD} atomic line database \citep{Kupka1999}, which uses data from \citet{Kurucz1993}, was used to 
create a line list for the synthetic spectrum.  Synthetic spectra were
generated using the {\tt SYNSPEC} code \citep{Hubeny1994,Krticka1998}.  We adopted
a microturbulence of 2\,km\,s$^{-1}$.  The computed spectrum was convolved with 
the instrumental profile (Gaussian of 0.4\,\AA ~FWHM for the Coud\`e spectra 
and 1.5\,\AA ~FWHM for the WIRO spectra respectively) and rotationally broadened 
to fit the observed spectrum under the assumption of uniform rotation.

\begin{figure}
\centering
\includegraphics[]{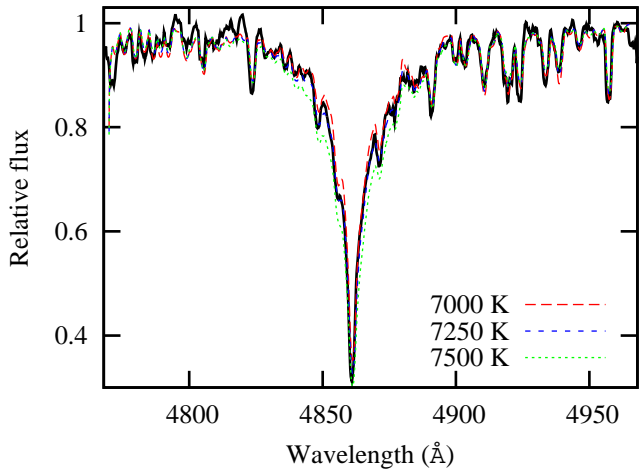}
\caption{The H$\beta$ line  fitted with a model with $T_{\rm eff}=7250~{\rm K}$, 
$\log g=3.5$ (cgs) (short dashes). Two other models with 
$T_{\rm eff}=7000~{\rm K}$ (long dashes) and $T_{\rm eff}=7500~{\rm K}$ 
(dots) are given for comparison.}
\label{fsyn}  
\end{figure}

The best fit to the H$\beta$ and H$\alpha$ lines was obtained for effective
temperature $T_{\rm eff}=7250~{\rm K}$, and surface gravity $\log g=3.5$. We used
the Mg\,II $\lambda$\,4481\,{\AA} line to determine the projected rotational velocity. 
The best match was obtained with $v\,\sin\,i=90~{\rm km\,s^{-1}}$. In Fig. \ref{fsyn} 
we show the best fit to H$\beta$ as well as the fit using two other models with 
different temperatures for comparison. Our results for the atmospheric parameters and 
$v\sin\,i$ are in very good agreement with the results given by
\citet{Balona2012a} who find $T_{\rm eff} = 7300 \pm 150$\,K, $\log g = 3.5
\pm 0.1$, $v \sin i = 100 \pm 10$\,km\,s$^{-1}$. \citet{Uytterhoeven2011}
list the effective temperature as $T_{\rm eff} = 6810 \pm 130$~K, which is
in disagreement with the above values.

\section{Photometry}
The photometric data were used to determine the amplitude-colour relation and hence
the mode for each frequency.The multi-site photometric campaign on V2367~Cyg was 
carried out during 2011 June to 2011 September (Julian dates 2455726.69 -- 2455816.52) 
in Europe and North America with five different sites. Although time allocation for
10~d at KISO Observatory, Japan was allocated, no photometric data could be 
obtained due to bad weather.  The time of these observations coincide with 
{\it Kepler} quarter Q9.3--Q10.3.  There are over 127780 short-cadence (1~min)
{\it Kepler} data points in this time range.  Table\,\ref{tab:log} summarizes 
the sites, telescopes and additional information regarding the campaign. 

Since the aim of this campaign is mode identification, multi-colour 
${\rm UBVRI}$ filters were chosen.  However, EUO and OAVdA do not have the 
U filter.  The U filter, when it is used, leads to high signal-to-noise
owing to the poor sensitivity of the CCDs in this wavelength and is
therefore not reliable.  In any case only four nights of $U$ photometry are
available which is not sufficient to resolve the frequencies.  Although we
show the $U$ observations in mode identification, they should be given very
low weight or ignored.  All R and I filters are of the Johnson/Bessel type
except those at SPM (Mexico) which are of the Cousins/Kron type.  The
difference is of little consequence as it has a negligible effect on the
amplitude ratio and phase difference which are used for mode identification.

When the campaign was initiated in 2011 June, all sites observed the target
for 3--4 nights.  Subsequently, only EUO continued to observe the target in 
July, August and early September.  A total of 181 hours were spent on
observing V2367\,Cyg within the 98-d duration of the campaign.  Data were 
reduced using standard {\tt IRAF} routines including correction for dark,
bias and flat-fielding of each CCD frame.  The {\tt DAOPHOT II} package
\citep{Stetson1987} was used for aperture photometry.  The observing times
were converted to heliocentric Julian Date (HJD). 

Since {\it Kepler} photometry was available, it was easy to check the
variability of nearby stars.  We used the comparison stars GSC\,3556 \,1047
(C1), GSC\,3556\,1343 (C2) and GSC\,3556\,627 (C3) because they showed least 
variability in the {\it Kepler} photometry (not more than 50 ppm).  Because 
not all instruments have the same field of view, C2 was not observed by some 
sites and C1 was chosen as the primary comparison star. The data was
therefore reduced relative to C1. Outliers produced by poor seeing or 
transparency variations were removed by visual inspection.

The passbands of corresponding filters in the instruments on different sites 
are not exactly the same and neither are the optical characteristics of the
instruments.  Therefore there are invariably zero-point differences in each
filter for each site.  Fortunately, we know the frequency composition of the
variations in  V2367\,Cyg from the analysis of the {\it Kepler} data by
\citet{Balona2012a}.  We can assume that the light variations are well
modeled by a Fourier series comprising these frequencies where the zero
point will vary from site to site.  This is easy to do by fitting the
following function to the data by least squares:
\begin{align}
y_i + \epsilon_i = \sum_{j=1}^s \delta_{ij} a_j + \sum_{k=1}^m A_k\sin(2\pi f_k t_i + \phi_k
\label{eq1})
\end{align}
where $y_i$ is the magnitude at time $t_i$ and $\epsilon_i$ is the unknown
observational error.  The number of different sites is $s$ and $\delta_{ij} =
0$ unless the $i$-th observation is from site $j$, in which case
$\delta_{ij} = 1$.  The last term is a truncated Fourier series with $m$ 
frequencies which represents the variability of  V2367\,Cyg.  By minimizing 
$\sum \epsilon^2_i$ for $1 \le i \le N$ ($N$ being the total number of 
observations), the individual zero points for each site, $a_j$, and the 
amplitudes and phases of the Fourier series can be found.

\begin{figure}
\centering
\includegraphics[]{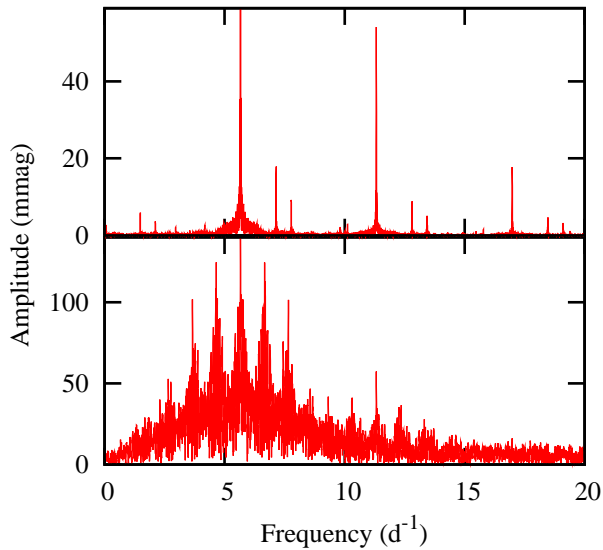}
\caption{Bottom panel: periodogram of the ground-based $V$ observations of 
V2367\,Cyg.  Top panel: periodogram of the {\it Kepler} observations over
the same time interval.}
\label{fig:pergm}
\end{figure}

In Fig.\,\ref{fig:pergm} we show the periodogram of the $V$ observations and
also of the {\it Kepler} observations over the same time interval
(JD~2455718.0--2455817.0).  The huge advantage of an uninterrupted time
series is very clear in this figure.

\begin{table}
\centering
\caption{Amplitudes, $A$ (mmag), and phases $\phi$ (radians) for the three
frequencies of highest amplitude: $f_1 = 5.66106$, $f_2 = 7.14895$, 
$f_3 = 7.77557$~d$^{-1}$ and harmonics of $f_1$.  The epoch of phase zero is
JD 2455700.000.}
\label{tab:coefs}
\begin{tabular}{lrr}
\hline
\multicolumn{1}{c}{Freq} & 
\multicolumn{1}{c}{$A$} &
\multicolumn{1}{c}{$\phi$} \\
\hline
       & \multicolumn{2}{c}{$U$}\\
$f_1$  & $  248.9 \pm 1.5$ & $ 2.3343 \pm 0.0060$ \\
$2f_1$ & $   77.5 \pm 1.4$ & $ 2.5586 \pm 0.0177$ \\
$f_2$  & $   24.0 \pm 1.4$ & $-0.5104 \pm 0.0586$ \\
$3f_1$ & $   24.7 \pm 1.4$ & $ 2.7732 \pm 0.0551$ \\
$f_3$  & $   11.9 \pm 1.5$ & $-0.7500 \pm 0.1249$ \\
$4f_1$ & $   12.5 \pm 1.4$ & $ 2.6230 \pm 0.1090$ \\
       & \multicolumn{2}{c}{$B$}\\
$f_1$  & $  243.4 \pm 1.1$ & $ 2.2941 \pm 0.0043$ \\
$2f_1$ & $   75.2 \pm 1.0$ & $ 2.4873 \pm 0.0140$ \\
$f_2$  & $   26.5 \pm 1.1$ & $-0.5388 \pm 0.0403$ \\
$3f_1$ & $   22.4 \pm 1.0$ & $ 2.7301 \pm 0.0467$ \\
$f_3$  & $   15.1 \pm 1.1$ & $-0.7942 \pm 0.0706$ \\
$4f_1$ & $   11.4 \pm 1.0$ & $ 2.7762 \pm 0.0918$ \\
       & \multicolumn{2}{c}{$V$}\\
$f_1$  & $  186.9 \pm 0.7$ & $ 2.2641 \pm 0.0035$ \\
$2f_1$ & $   55.1 \pm 0.6$ & $ 2.4618 \pm 0.0117$ \\
$f_2$  & $   19.4 \pm 0.7$ & $-0.4727 \pm 0.0351$ \\
$3f_1$ & $   16.0 \pm 0.6$ & $ 2.7474 \pm 0.0401$ \\
$f_3$  & $   10.8 \pm 0.7$ & $-0.7081 \pm 0.0623$ \\
$4f_1$ & $    7.9 \pm 0.6$ & $ 2.7543 \pm 0.0806$ \\
       & \multicolumn{2}{c}{$R$}\\
$f_1$  & $  141.7 \pm 0.7$ & $ 2.2406 \pm 0.0050$ \\
$2f_1$ & $   44.9 \pm 0.7$ & $ 2.4660 \pm 0.0155$ \\
$f_2$  & $   17.3 \pm 0.7$ & $-0.3476 \pm 0.0407$ \\
$3f_1$ & $   13.6 \pm 0.7$ & $ 2.8603 \pm 0.0516$ \\
$f_3$  & $    9.5 \pm 0.7$ & $-0.5988 \pm 0.0746$ \\
$4f_1$ & $    6.2 \pm 0.7$ & $ 2.6144 \pm 0.1124$ \\
       & \multicolumn{2}{c}{$I$}\\
$f_1$  & $  113.0 \pm 0.6$ & $ 2.1942 \pm 0.0055$ \\
$2f_1$ & $   36.2 \pm 0.6$ & $ 2.4433 \pm 0.0169$ \\
$f_2$  & $   13.7 \pm 0.6$ & $-0.4963 \pm 0.0449$ \\
$3f_1$ & $   10.6 \pm 0.6$ & $ 2.8294 \pm 0.0579$ \\
$f_3$  & $    8.3 \pm 0.6$ & $-0.6936 \pm 0.0746$ \\
$4f_1$ & $    5.3 \pm 0.6$ & $ 2.6109 \pm 0.1147$ \\
\\
\hline
\end{tabular}
\end{table}

\begin{table}
\centering
\caption{Amplitude ratios, $A/A_B$, and phase differences $\phi - \phi_B$ 
(radians) relative to the $B$ band for six frequencies of highest amplitude: 
$f_1 = 5.66106$, $f_2 = 7.14895$,  $f_3 = 7.77557$~d$^{-1}$.}
\label{tab:modeid}
\begin{tabular}{lrr}
\hline
\multicolumn{1}{c}{Filt} & 
\multicolumn{1}{c}{$A/A_B$} &
\multicolumn{1}{c}{$\phi-\phi_B$} \\
\hline
       & \multicolumn{2}{c}{$f_1$}\\
  $U$  & $ 1.0226 \pm 0.0077$ & $ 0.0402 \pm 0.0074$ \\
  $B$  & $ 1.0000 \pm 0.0064$ & $ 0.0000 \pm 0.0061$ \\
  $V$  & $ 0.7679 \pm 0.0045$ & $-0.0300 \pm 0.0055$ \\
  $R$  & $ 0.5822 \pm 0.0039$ & $-0.0535 \pm 0.0066$ \\
  $I$  & $ 0.4643 \pm 0.0032$ & $-0.0999 \pm 0.0070$ \\
           & \multicolumn{2}{c}{$2f_1$}\\
  $U$  & $1.0306 \pm 0.0231$ & $ 0.0713 \pm 0.0226$ \\
  $B$  & $1.0000 \pm 0.0188$ & $ 0.0000 \pm 0.0198$ \\
  $V$  & $0.7327 \pm 0.0126$ & $-0.0255 \pm 0.0182$ \\
  $R$  & $0.5971 \pm 0.0122$ & $-0.0213 \pm 0.0209$ \\
  $I$  & $0.4814 \pm 0.0102$ & $-0.0440 \pm 0.0219$ \\
       & \multicolumn{2}{c}{$f_2$}\\
  $U$  & $0.9057 \pm 0.0648$ & $ 0.0284 \pm 0.0711$ \\
  $B$  & $1.0000 \pm 0.0587$ & $ 0.0000 \pm 0.0570$ \\
  $V$  & $0.7321 \pm 0.0403$ & $ 0.0661 \pm 0.0534$ \\
  $R$  & $0.6528 \pm 0.0378$ & $ 0.1912 \pm 0.0573$ \\
  $I$  & $0.5170 \pm 0.0312$ & $ 0.0425 \pm 0.0603$ \\
       & \multicolumn{2}{c}{$3f_1$}\\
  $U$  & $1.1027 \pm 0.0796$ & $ 0.0431 \pm 0.0722$ \\
  $B$  & $1.0000 \pm 0.0631$ & $ 0.0000 \pm 0.0660$ \\
  $V$  & $0.7143 \pm 0.0416$ & $ 0.0173 \pm 0.0616$ \\
  $R$  & $0.6071 \pm 0.0414$ & $ 0.1302 \pm 0.0696$ \\
  $I$  & $0.4732 \pm 0.0341$ & $ 0.0993 \pm 0.0744$ \\
       & \multicolumn{2}{c}{$f_3$}\\
  $U$  & $0.7881 \pm 0.1147$ & $ 0.0442 \pm 0.1435$ \\
  $B$  & $1.0000 \pm 0.1030$ & $ 0.0000 \pm 0.0998$ \\
  $V$  & $0.7152 \pm 0.0697$ & $ 0.0861 \pm 0.0942$ \\
  $R$  & $0.6291 \pm 0.0652$ & $ 0.1954 \pm 0.1027$ \\
  $I$  & $0.5497 \pm 0.0564$ & $ 0.1006 \pm 0.1027$ \\
       & \multicolumn{2}{c}{$4f_1$}\\
  $U$  & $1.0965 \pm 0.1560$ & $-0.1532 \pm 0.1425$ \\
  $B$  & $1.0000 \pm 0.1241$ & $ 0.0000 \pm 0.1298$ \\
  $V$  & $0.6930 \pm 0.0804$ & $-0.0219 \pm 0.1222$ \\
  $R$  & $0.5439 \pm 0.0778$ & $-0.1618 \pm 0.1451$ \\
  $I$  & $0.4649 \pm 0.0666$ & $-0.1653 \pm 0.1469$ \\
\\
\hline
\end{tabular}
\end{table}

\begin{table}
\centering
\caption{Zero points in the ${\rm UBVRI}$ filters for various sites and the number
of observations, $N$.}
\label{tab:zeros}
\begin{tabular}{lrrrr}
\hline
\multicolumn{1}{c}{Site} & 
\multicolumn{1}{c}{$U$} &
\multicolumn{1}{c}{$N$} \\
\hline
Mexico     & $-0.078 \pm 0.001$ & $  448$ \\
USA     & $-0.033 \pm 0.003$ & $   44$ \\
\hline
\multicolumn{1}{c}{Site} & 
\multicolumn{1}{c}{$B$} &
\multicolumn{1}{c}{$N$} &
\multicolumn{1}{c}{$V$} &
\multicolumn{1}{c}{$N$} \\
\hline
Bulgaria     & $-0.193 \pm 0.003$ & $  220$ & $-0.079 \pm 0.002$ & $  221$\\
Italy     & $-0.210 \pm 0.002$ & $  372$ & $-0.120 \pm 0.002$ & $  412$\\
Mexico     & $-0.170 \pm 0.001$ & $ 1319$ & $-0.081 \pm 0.001$ & $ 2152$\\
Turkey     & $-0.205 \pm 0.001$ & $ 1955$ & $-0.121 \pm 0.001$ & $ 2128$\\
USA     & $-0.170 \pm 0.007$ & $   50$ & $-0.066 \pm 0.005$ & $   49$\\
\hline
\multicolumn{1}{c}{Site} & 
\multicolumn{1}{c}{$R$} &
\multicolumn{1}{c}{$N$} &
\multicolumn{1}{c}{$I$} &
\multicolumn{1}{c}{$N$} \\
\hline
Bulgaria     & $-0.007 \pm 0.002$ & $  220$  & $ 0.075 \pm 0.002$ & $  220$ \\
Italy     & $-0.037 \pm 0.001$ & $  410$  & $ 0.056 \pm 0.001$ & $  399$ \\
Mexico    & $ 0.091 \pm 0.001$ & $  395$  & $ 0.092 \pm 0.001$ & $  375$ \\
Turkey     & $-0.057 \pm 0.001$ & $ 2052$  & $-0.002 \pm 0.001$ & $ 2008$ \\
USA     & $ 0.013 \pm 0.004$ & $   50$  & $ 0.092 \pm 0.004$ & $   46$ \\
\\                                     
\hline
\end{tabular}
\end{table}

\begin{figure*}
\centering
\includegraphics[]{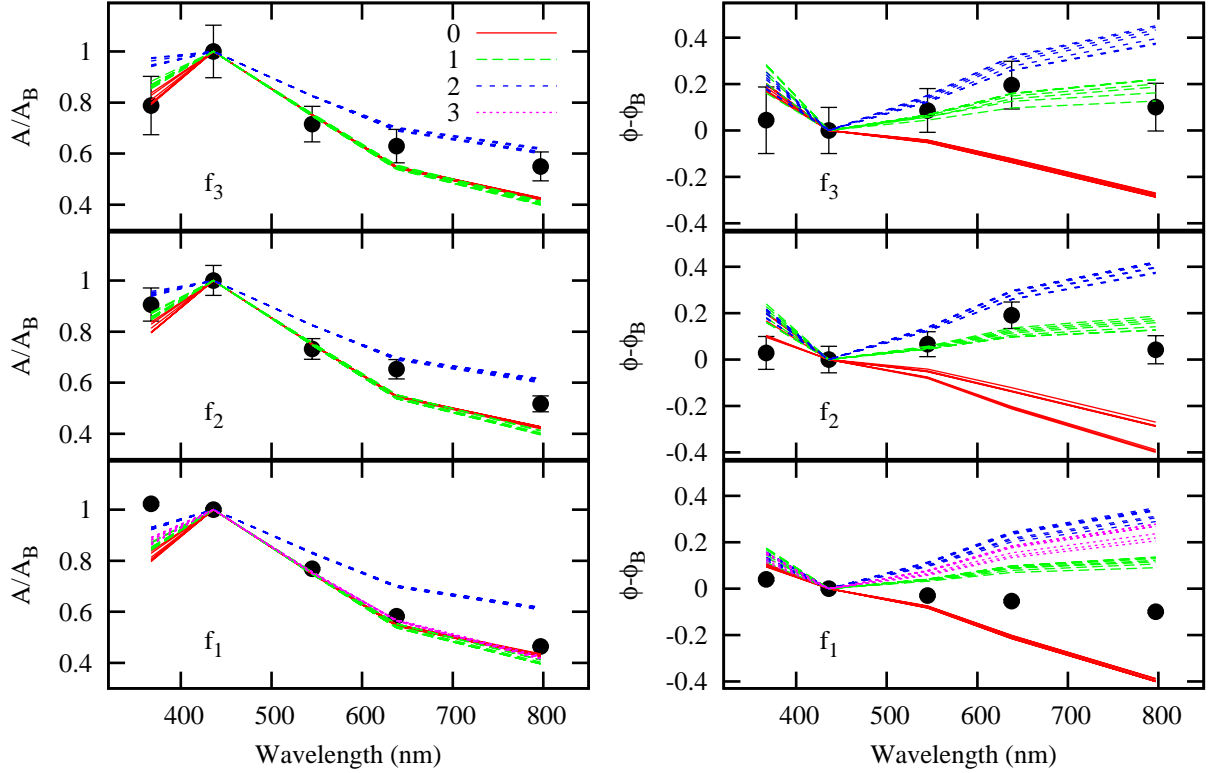}
\caption{Amplitude ratios (left panels) and phase differences (radians, right panels)
for $f_1, f_2$ and $f_3$.  The curves are from models with a range of stellar
parameters of the best estimated values with a range of stellar parameters of $7200 < T_{\rm eff} < 7400$~K and $3.3 < \log g < 3.7$, and for different spherical harmonic
degrees, $l$($0<l<3$).}
\label{fig:modeid}
\end{figure*}

\begin{figure*}
\centering
\includegraphics[]{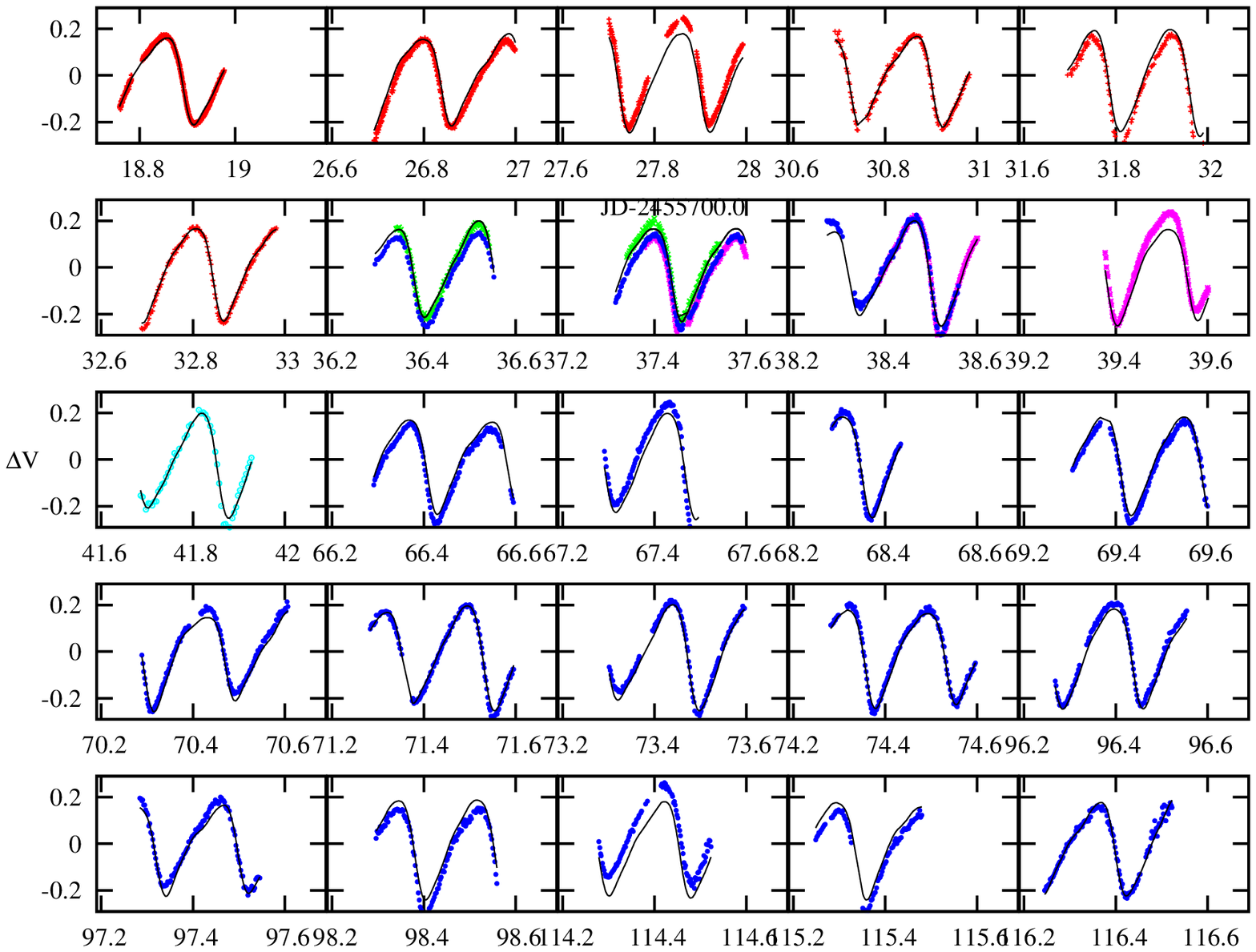}
\caption{Light curves corrected for zero points showing fitted Fourier
curve.}
\label{fig:lcurve}
\end{figure*}

We used the first three independent frequencies of highest amplitude obtained 
by the frequency extraction described in \citet{Balona2012a}: 
$f_1 = 5.661058$, $f_2 = 7.148949$ and $f_3 = 7.775566$\,d$^{-1}$ together
with the first three harmonics of $f_1$ ($2f_1, 3f_1, 4f_1$) for the least
squares Fourier fit (Eq.\ref{eq1}).  Fitting with these six frequencies does not 
adequately describe the full variation of the star, but is sufficient to 
determine the zero points of different sites (Table\,\ref{tab:zeros})
together with the amplitudes and phases of these particular six frequencies 
(Table\,\ref{tab:coefs}).  It is these amplitudes and phases which are 
used in mode identification.  Differential $V$-band photometry from all sites 
adjusted for zero-point differences are shown in Fig.\,\ref{fig:lcurve}.  
Also shown is the fitted Fourier curve using the six frequencies.
From the amplitudes and phases of the fitted Fourier frequency components, 
we find amplitude ratios and phase differences listed in Table.\,\ref{tab:modeid} 
and shown in Fig.\,\ref{fig:modeid}.

\begin{figure*}
\centering
\includegraphics[]{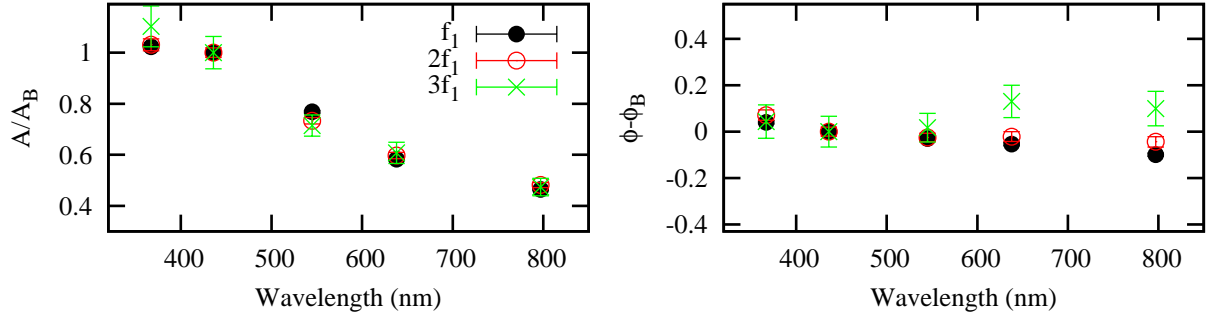}
\caption{Amplitude ratios (left panel) and phase differences (radians, right 
panel) for $f_1$ and its harmonics, $2f_1, 3f_1$.}
\label{fig:f1harm}
\end{figure*}

The harmonics of the principal frequency, $f_1$, have large amplitudes and
can be used as additional data for mode identification.  In fact, the
amplitude ratios and phase differences for the harmonics $2f_1, 3f_1$, are 
exactly the same as for the fundamental, $f_1$, as can be seen in
Fig.\,\ref{fig:f1harm}.

\section{Mode identification}

There are several techniques that can be used to discriminate between
different values of the spherical harmonic degree, $l$.  In the simplest 
technique, the amplitudes ratios relative to a given waveband are compared with
the calculated amplitudes ratios.  The method relies on the fact that the 
variation of light amplitude in different wavelength bands depends on $l$ and
is independent of the azimuthal number $m$. It should be noted that our current 
models for photometric mode identification are only valid in the limit of no rotation.
Rotation introduces departures from pure spherical harmonics so that a
single value of $l$ is not sufficient to describe the eigenfunction
perfectly.  However, so long as the rotation is not very rapid, a single
value of $l$ is still a reasonable approximation.  We expect that for V2367
Cyg this will be the case, but the relative amplitudes will be modified to
some degree due to this effect.

The most general expression for the magnitude variation, $m_\lambda$,
at wavelength $\lambda$ for a star pulsating with spherical harmonic
degree, $l$, and angular frequency of pulsation, $\omega$, with
angle of inclination, $i$, is: 
\begin{align*} \Delta m_\lambda  &= -\frac{2.5}{\ln 10} \epsilon P_l^ m(\cos i) b_{l \lambda}
\left\{ (1-l)(l+2)\cos(\omega t)\right .\\ 
&\left .+ f_T(\alpha_{T\lambda} + \beta_{T\lambda})\cos(\omega t + \psi_T)
 - f_g(\alpha_{g\lambda} + \beta_{g\lambda}) \right \}
\end{align*} 
where $\epsilon$ is the amplitude of the radial displacement, $P_l^m$ is the
associated Legendre polynomial, $f_T = \delta T_{\rm eff}/T_{\rm eff}$
is the amplitude of local effective temperature variation, $f_g = \delta
g_e/g_e$ is the amplitude of the local gravity variation for unit radial
displacement at the photosphere and $\psi_T$ is the phase difference
between effective temperature variation and radial displacement.
The quantity $b_{l\lambda} = \int_0^1 \mu h_\lambda P_l(\mu) d\mu$
where the normalized limb darkening function is given by $h_\lambda$
and $\mu = \cos \theta$ is the polar angle in the spherical coordinate
system where the $z$-axis is the line of sight to the observer.

The first term within the bracket in the above expression represents
the change in magnitude caused by the geometrical variation of the star
during pulsation and is independent of wavelength.  The other two terms
are wavelength dependent and lead to a variation of amplitude and phase
with waveband.  The differences in amplitude and phase at different
wavebands are small and accurate multicolour photometry is required for
good discrimination in $l$.  Mode discrimination is improved by using
wavebands covering as wide a wavelength range as possible.

The quantities $\alpha_{T\lambda}$, $\beta_{T\lambda}$,
$\alpha_{g\lambda}$, $\beta_{g\lambda}$ are defined in 
\citet{Balona1999} and can be determined from static model atmospheres.
The assumption that at any instant the atmosphere of a pulsating star
has the same structure as a static star of the same effective temperature
and gravity, is probably a reasonable one if the frequency of pulsation
is low. At any rate this assumption is required in order to make progress
with the current state of the art.
The quantity, $f_T$, the amplitude of local effective temperature
variation relative to radial displacement, is an important parameter
in mode identification.  It turns out that the calculation of $f_T$
is sensitive to the treatment of convection in $\delta$~Sct models
\citep{Balona1999}. One way of overcoming this difficulty is to leave $f_T$
as a free parameter to be determined by the observations using a $\chi^2$
minimization technique \citep{Daszynska2003}. Application of this method
to $\delta$~Sct stars indicates that the simple mixing-length theory for
convection is inadequate, since the observed values of $f_T$ are not in
agreement with the calculated values.

In mode identification we therefore have the option of leaving $f_T$ as a
free parameter to be determined by the observations themselves or to use the
somewhat inadequate models to constrain the solution. We decided to use  
the latter option because leaving $f_T$ free increases the number of
parameters by two. We took the view that the effect of rotation, in any
case, would limit the accuracy of the results so there is nothing to be
gained by leaving $f_T$ as a free parameter. The uncertainty in $f_T$
means, however, that not much importance can be placed in the resulting
phase differences from the models. The amplitude of $f_T$ is more reliable
and we can place more confidence in the calculated amplitude ratios.

Balona et al. (2012) estimated the effective temperature $T_{\rm eff} =
7300 \pm 150$~K and luminosity $\log L/L_\odot = 1.7 \pm 0.1$, which places
the star roughly in the middle of the instability strip and at the end of
core hydrogen burning.  There is a thin convection zone of H and He I near
the surface which might modify the phase of the eigenfunction, but our
current understanding of convection is insufficient to calculate this effect
with any certainty.  We may thus expect good agreement with the models for
the relative amplitudes in various wavebands but less good agreement for the
relative phases.

The calculation of amplitude ratios and phase differences was performed using
the {\tt FAMIAS} software package \citep{Zima2008}. The photometric module of 
the software package compares the observed parameters with a precomputed grid
of non-adiabatic models for a range of stellar parameters.  The program
produces amplitude ratios and phase differences in a number of different
passbands which include the Geneva, Johnson/Cousins and Str\"{o}mgren
systems.  In our case we used the results from the Johnson/Cousins 
system normalized to the $B$ filter.

Since the amplitude ratios depend to some extent on the stellar parameters,
it is important to calculate these values over a sufficiently wide range of
models centered on the best estimate of the parameters which we take to be:
mass $M/M_\odot = 2.10$, effective temperature $T_{\rm eff} = 7300$\,K and 
$\log g = 3.5$.   We calculated amplitude ratios and phase differences for 
models in the range $7200 < T_{\rm eff} < 7400$~K and $3.3 < \log g < 3.7$.  
These models were used because they produced the closest match to the observed
frequencies $f_1, f_2$ and $f_3$.  In {\tt FAMIAS}, models for $\delta$~Sct 
stars computed by the {\tt ATON} code \citep{Ventura2008} and {\tt MAD} 
\citep{Montalban2007}.  The results given by {\tt FAMIAS} for different 
values of $l$ are compared with the observed values in Fig.\,\ref{fig:modeid}.

\section{Discussion}

At moderate rotation rates and for close frequencies, a strong coupling
exists between modes with $l$ differing by 2 and of the same azimuthal order, 
$m$. A consequence of mode coupling is that modes of higher degree should be
considered in photometric mode identification. Modes with nominal degree
$l>2$ acquire substantial $l \le 2$ components and therefore are
more likely to reach detectable amplitudes.  In these circumstances,
mode identification using amplitude ratios or phase differences becomes
both aspect- and $m$-dependent \citep{Daszynska2002}.

V2367\,Cyg is a surprisingly rapid rotator.  It can therefore be
expected that problems involving mode coupling will be significant and that
non-adiabatic quantities calculated from non-rotating models will not be
very reliable.  In particular, we expect that the observed amplitude ratios
for $f_2$ will be intermediate between $l = 0$ and $l = 2$ if mode coupling
is indeed present in this mode as suggested by \citet{Balona2012a}.

From Fig\,\ref{fig:modeid}, we see that $f_1$ could be $l = 0, 1$ or 3. 
Unfortunately in this range of stellar parameters there is no discrimination
between these three spherical harmonic degrees in the amplitude ratio.  The
phase differences are intermediate between $l = 0$ and $l = 1$.  What we can
conclude is that our results for $f_1$ are certainly consistent with $l =
0$ and exclude $l = 2$.

For $f_2$ the amplitude ratios are intermediate between $l = 0$ and $l = 2$,
as might be expected if this is a coupled mode.  The phase differences are
in agreement, though clearly no definite conclusion can be made on this
basis alone.  We conclude that mode coupling, as proposed by
\citet{Balona2012a}, is consistent with our results.  Results for $f_3$ are 
less reliable owing to the smaller amplitude, but are very similar to $f_2$, 
so this might again be a coupled mode.

Our overall conclusion from multicolour photometry is that mode
identification is in agreement with the identifications proposed by
\citet{Balona2012a}, but that the complexity introduced by the rapid rotation
of V2367\,Cyg produces severe limitations on any interpretation.  We
certainly cannot find any evidence to contradict the identifications, which
in itself is a useful result.

Although HADS stars are of particular interest as they offer insights into
nonlinear phenomena, the class itself is very poorly defined.  There is no
particular reason why the pulsation amplitude should be used as a criterion. 
Perhaps a better definition of the class is the presence of harmonics of the
dominant frequency or of combination frequencies.  It would also be very
important to refine the physical parameters of HADS.  If they are indeed
intermediate between $\delta$~Sct stars and Cepheids, they should line in a
region of the HR diagram between these two classes of stars.  More precise
effective temperatures and luminosities are required to test this
hypothesis.

\section*{Acknowledgments}
CU sincerely thanks the South African National Research Foundation for the 
prize of innovation post doctoral fellowship with the grant number 73446.
TG would like to thank NRF Equipment-Related Mobility Grant-2011 for travel to Turkey 
to carry out the photometric observations.
IS and II gratefully acknowledge the partial support from Bulgarian NSF under 
grant DO 02-85. DD acknowledges for the support of grants DO 02-362 and DDVU 
02/40-2010 of Bulgarian NSF.  
LAB thanks the South African National Research Foundation and the South 
African Astronomical Observatory for generous financial support. 
HAK acknowledges Carlos Vargas-Alvarez, Michael J.
Lundquist, Garrett Long, Jessie C. Runnoe, Earl S. Wood, Michael J.
Alexander for helping with the observations at WIRO.
BY wishes to thank EUO for the allocation time of observations during the campaign.
LFM acknowledges financial support from the UNAM under grant PAPIIT
IN104612 and from CONACyT by way of grant CC-118611.
MD, AC and DC are supported by grants provided by the European Union, the 
Autonomous Region of the Aosta Valley and the Italian Department for Work, 
Health and Pensions. The OAVdA is supported by the Regional Government of the 
Aosta Valley, the Town Municipality of Nus and the Mont Emilius Community.
TEP aknowledges support from the National Research Foundation of South Africa.
This study made use of {\tt IRAF} Data Reduction and Analysis System and the 
Vienna Atomic Line Data Base (VALD) services.  The authors thank Dr Zima for
providing the {\it FAMIAS} code.

\bibliographystyle{mn2e}
\bibliography{ceren}

\label{lastpage}

\end{document}